\documentclass[a4paper, amsfonts, amssymb, amsmath, reprint, showkeys, nofootinbib, twoside]{revtex4-1}
\usepackage[english]{babel}
\usepackage[utf8]{inputenc}
\usepackage[colorinlistoftodos, color=green!40, prependcaption]{todonotes}
\usepackage{amsthm}
\usepackage{amsmath}
\usepackage{braket}
\usepackage{mathtools}
\usepackage{physics}
\usepackage{xcolor}
\usepackage{graphicx}
\usepackage[left=23mm,right=13mm,top=35mm,columnsep=15pt]{geometry} 
\usepackage{adjustbox}
\usepackage{placeins}
\usepackage[T1]{fontenc}
\usepackage{lipsum}
\usepackage{csquotes}
\usepackage{verbatim}
\usepackage[pdftex, pdftitle={Article}, pdfauthor={Author}]{hyperref}
\usepackage{color,soul}
\usepackage{comment}
\usepackage{ulem}

\newcommand{\e}[1]{\textrm{e}^{#1}}
\newcommand{\dint}{\textrm{d}} 
\newcommand{\be}{\begin{eqnarray}}
\newcommand{\ee}{\end{eqnarray}}

\begin{document}
\title{Quantum Brownian Motion in the Caldeira-Leggett Model\\with a Damped Environment}

\author{Lester Buxton\textsuperscript{1,2}}
\author{Marc-Thomas Russo\textsuperscript{1,2}}
\author{Jim Al-Khalili\textsuperscript{1}}
\author{Andrea Rocco\textsuperscript{1,3}}
\email[Correspondence email address: ]{a.rocco@surrey.ac.uk}

\affiliation{\textsuperscript{1}School of Mathematics and Physics, \textsuperscript{2}Leverhulme Quantum Biology Doctoral Training Centre and \textsuperscript{3}Department of Microbial Sciences, University of Surrey, GU2 7XH, Guildford, UK}

\hspace*{0.5cm}

\date{\today} 

\begin{abstract}
We model a quantum system coupled to an environment of damped harmonic oscillators by following the approach of Caldeira-Leggett and adopting the Caldirola-Kanai Lagrangian for the bath oscillators. In deriving the master equation of the quantum system of interest (a particle in a general potential), we show that the potential is modified non-trivially by a new inverted harmonic oscillator term, induced by the damping of the bath oscillators. We analyze numerically the case of a particle in a double-well potential, and find that this modification changes both the rate of decoherence at short times and the well-transfer probability at longer times. We also identify a simple rescaling condition that keeps the potential fixed despite changes in the environmental damping. Here, the increase of environmental damping leads to a slowing of decoherence.
\end{abstract}

\keywords{Open quantum systems, Dissipation, Caldirola-Kanai Lagrangian}

\maketitle

\section{Introduction} \label{sec:Intro}

Dissipative dynamics cannot be derived from time-independent Hamiltonians or Lagrangians. While classical equations of motion can be defined so as to include phenomenological dissipation, in quantum mechanics the lack of  Hamiltonian or Lagrangian functions based on first principles makes quantum dissipative systems hard to describe.

Historically, this problem has been regarded as a major conceptual obstacle to the quantization of dissipative systems, and has sparked a considerable effort to identify a dynamical derivation of dissipative features. The celebrated work of Caldeira and Leggett \cite{Caldeira83} tackles this issue by defining a microscopic model of dissipation based on the interaction of a system of interest with its own environment, defined as a thermal bath of harmonic oscillators. While the elimination of the environmental degrees of freedom by projection leads to a dissipative system described macroscopically by the Langevin equation, the original system, before the reduction, is fully Hamiltonian, and as such can be quantized with standard procedures.  

The approach of \cite{Caldeira83} has been extended to consider both Markovian and non-Markovian dynamics \cite{Lally22}, different types of systems of interest, different structured environments \cite{Gribben20}, with repercussions in the broad fields of quantum technologies, quantum optics and quantum computing, and more generally in all those applications where controlling the behaviour of a quantum system coupled to its environment is necessary \cite{Schlosshauer04}.    

In this paper we further extend the approach of \cite{Caldeira83} to the case where the quantum environment is itself experiencing dissipative processes.

Our work follows in spirit the approach presented in \cite{Plyukhin19} in the classical case, where classical bath oscillators undergo dissipation, described by equations of motion with friction terms. However, in contrast to \cite{Plyukhin19}, in the quantum case we need to define an appropriate phenomenological Lagrangian formalism in order to be able to adopt the path integral approach defined in \cite{Caldeira83}. The solution to this problem is provided by considering the proposal by Caldirola \cite{Caldirola41} and Kanai \cite{Kanai48}, who introduced a method to describe dissipation using time-dependent Hamiltonians and Lagrangians, obtained by assuming the mass of the system to increase exponentially in time \cite{Crespo90,Choi13}. While this approach has been received with some degree of controversy \cite{Dekker81,Hensel92}, it has nonetheless been used widely to describe dissipative systems \cite{Baskoutas94,Dodonov07,Tokieda17,Tokieda21}.

Here, we use the propagator associated with a damped oscillator described by the Caldirola-Kanai Lagrangian \cite{Kerner58,Um87} to implement damping on the bath oscillators in the Caldeira-Leggett model. The damped environment so defined can still be integrated out analytically and the corresponding influence functional derived. This leads to a master equation for the system of interest in which the potential experienced by the system is modified by a new damping-dependent term which takes the form of an inverted harmonic potential.

We study the time evolution and dynamics of a particle of mass $m$ in a symmetric double well potential by focusing on two relevant processes: (1) the decoherence process at short times for the particle initially in a superposition of identical Gaussian wave functions, one in each well, and (2) the dynamics of the particle at long times, starting as a single Gaussian distribution on one side of the well. We then explore the time-dependence of the subtle interplay between two parameters: the strength of the damping felt by the environment oscillators and the relaxation rate of the system of interest (the particle). For the case of the particle starting on one side of the barrier, we explore transitions between the regimes of parameter space in which well-transfer (whether via tunnelling through the barrier or classical over-the-barrier hopping) is reduced (a quantum Zeno effect \cite{misra}) or enhanced due to thermal excitation by the environment (an anti-Zeno effect) \cite{Kofman00,Facchi01}. Furthermore, by requiring that the modified potential felt by the system is invariant upon rescaling of the environmental damping, we find that dissipation decreases and decoherence slows down as the environmental damping is increased.

In Sec.\ref{sec:Model} we introduce the Caldirola-Kanai Lagrangian into the Caldeira-Leggett model to describe an environment of damped harmonic oscillators, and derive the reduced master equation for the system of interest. In Sec.\ref{sec:Dynamics} we discuss the dynamics that emerge from our model, focusing in particular on decoherence and well-transfer processes, and in Sec.\ref{sec:Conclusions} we present some concluding remarks. 

\section{The Caldeira-Leggett model with damped bath oscillators} \label{sec:Model}

In order to introduce dissipation phenomenologically in the bath oscillators, we follow the approach of  Caldirola and Kanai \cite{Caldirola41,Kanai48}  for a classical damped harmonic oscillator.  The Lagrangian of a standard harmonic oscillator without damping is given by 
\begin{equation}\label{eq:HOLag}
    \mathcal{L}(R,\dot{R})=\frac{1}{2}m\dot{R}^2  -  \frac{1}{2}m\omega^2R^2,
\end{equation} 
where $R$ and $\dot{R}$ are the coordinate and velocity of the oscillator, respectively, $m$ is its mass, and $\omega$ its frequency. As in \cite{Caldirola41,Kanai48}, we extend this Lagrangian to become time-dependent by assuming that the mass of the oscillator grows exponentially in time with rate $\mu$:
\begin{equation}
    m(t) = m_0 \e{\mu t}. \label{mexp}
\end{equation}
This results in the Caldirola-Kanai Lagrangian \cite{Caldirola41,Kanai48},
\begin{equation}\label{eq:CKLag}
    \mathcal{L}(R,\dot{R},t) = \e{\mu t}\left(\frac{1}{2}m_0\dot{R}^2 - \frac{1}{2}m_0\omega^2R^2\right),
\end{equation}
which leads to the desired equation of motion of the form
\begin{equation}
    \ddot{R}(t)+\mu\dot{R}(t)+\omega^2R(t)=0. \label{hof}
\end{equation}
The friction term $\mu \dot{R}$ in Eq. (\ref{hof}) emerges as a result of the assumption in Eq. (\ref{mexp}) and describes a damped harmonic oscillator. 

This Lagrangian can now be used to describe the bath oscillators in the quantum Brownian motion model of Caldeira and Leggett \cite{Caldeira83}. We begin by defining the Lagrangian for the system of interest coupled to a heat bath of oscillators as:
 
\begin{equation}
    \mathcal{L}= \mathcal{L}_A + \mathcal{L}_B + \mathcal{L}_I. \label{totlag}
\end{equation}

Here, $\mathcal{L}_A$ is the Lagrangian of the system of interest, characterised by spatial coordinate $x$ and a generic potential $V(x)$:  
\begin{equation}\label{eq:LS}
    \mathcal{L}_A=\frac{1}{2}M\dot{x}^2(t)-V(x).
\end{equation}

The term $\mathcal{L}_B$ in Eq. (\ref{totlag}) represents the bath oscillators, assumed to experience a damping process described by the Caldirola-Kanai Lagrangian  
\begin{equation}\label{eq:LB}
    \mathcal{L}_B= \sum_k\e{\mu_k t}\left(\frac{1}{2}m_k\dot{R}_k^2 - \frac{1}{2}m_k\omega_{0_k}^2R_k^2\right).
\end{equation}
For the sake of simplicity we will assume from now on that $\mu_k \equiv \mu$ for all $k$'s. This assumption, despite being restrictive, is consistent with the view of the bath oscillators undergoing a damping process induced on each of them by a common secondary environment, as analyzed for instance in \cite{Plyukhin19}.  

Finally, the term $\mathcal{L}_I$ in Eq. (\ref{totlag}) is the interaction Lagrangian between system and bath, with coupling constant $C_k$,
\begin{equation}\label{eq:LI}
    \mathcal{L}_I=-x \sum_kC_kR_k .
\end{equation}

By following the standard procedure \cite{Caldeira83,Feynman63,Feynman65}, we construct the evolution equation for the reduced density matrix of the system, $\Tilde{\rho}$, namely:
\begin{equation}
    \Tilde{\rho}(x,y,t) = \int dx'dy' J(x,y,t;x',y',0)\rho_A(x',y',0). \label{rhoprop}
\end{equation}
Here the superpropagator $J(x,y,t;x',y',0)$ reads
\begin{equation}
\begin{aligned}\label{eq:OurSuperpropagaotr}
    J(x,y,t;x',y',0) =& \iint \mathcal{D}x\mathcal{D}y \mathcal{F}[x,y]\e{\frac{i}{\hbar}(S_A[x]-S_A[y])},
    \end{aligned}
\end{equation}
with the influence functional $\mathcal{F}[x,y]$ given by
\be
&&\mathcal{F}[x,y] = \int dR^\prime dQ^\prime dR \rho_B(R^\prime,Q^\prime,0) \nonumber \\
&& \qquad\qquad\qquad \times K(x,R,R^\prime) K^*(y,Q,Q^\prime)\ . \label{infl}
\ee

In Eq. (\ref{rhoprop}) and Eq. (\ref{infl}), the total density matrix is assumed to be factorised at time zero as $\rho_T(0)=\rho_A(0) \rho_B(0)$, and $K(x,R,R^\prime)$ represents the propagator of the bath oscillators, damped as described by the Lagrangian, Eq. (\ref{eq:LB}), and driven by the system of interest through the interaction, Eq. (\ref{eq:LI}). The initial coordinates at time 0 are denoted by a prime, so that $x(0)=x'$, whereas the coordinates at time $t$ are not, so that $x(t)=x$.

In order to compute the propagator $K(x,R,R^\prime)$, we make use of the results of \cite{Um87}, who obtained the propagator of the damped and forced quantum harmonic oscillator 
\begin{equation}
m_0 \ddot{R}(t)+m_0 \mu\dot{R}(t)+m_0 \omega^2R(t)=F(t), \label{eqmU}
\end{equation}
by using Feynman's path integral formalism \cite{Feynman65}. To adapt the results of \cite{Um87} to our case, it suffices to  replace the external forcing term $F(t)$ in Eq. (\ref{eqmU}) with $F(t)=x(t)\e{-\mu t}\sum_kC_k$. Hence, the propagator for the bath oscillators results in
\begin{multline}
  K(x,R,R') = A(t) \exp\Big\{\frac{i}{\hbar}\Big(a(t) R^2+b(t)R'^2\\
  +c(t) R R'+d(t)R+e(t)R'+g(t)\Big)\Big\}, \label{prpdamp}
\end{multline}
where the pre-factor, $A(t)$, is given by the expression
\begin{equation}
    A(t) = \left(\frac{m\omega \e{\mu t/2}}{2\pi i \hbar \sin \omega t}\right)^{1/2},
\end{equation}
and the time-dependent coefficients in Eq. (\ref{prpdamp}) are
\be
&&a(t)=\frac{m}{4}(-\mu+2\omega \cot\omega t) \e{\mu t},\\
&&b(t)= \frac{m}{4}(\mu+2\omega \cot\omega t), \\
&&c(t) = -\frac{m\omega}{\sin\omega t}\e{\mu t/2}, \\
&&d(t) = \frac{e^{\mu t/2}}{\sin(\omega t)} \int^t_0x(t')e^{-\mu t'/2} \sin(\omega t') \dint t' ,\\
&&e(t) =  \frac{1}{\sin(\omega t)} \int^t_0 x(t') e^{-\mu t'/2}\sin(w(t-t'))\dint t',\\
&&g(t)=- \frac{1}{m\omega\sin(\omega t)}\int^t_0\int^{t'}_0 x(t')x(s)\nonumber \\
&&\qquad e^{-\mu(t'+s)/2} \sin(w(t-t')) \sin(ws) \dint s \dint t'.
\ee

The influence functional is therefore:
\begin{multline}
    \mathcal{F}[x,y] =\exp\bigg\{-\frac{1}{\hbar}\bigg[\int^{t}_{0} f_{-}(t')\e{-\frac{\mu}{2} t'}I_{I}(t')dt'\\ 
    +\int^{t}_{0} f_{-}(t')\e{-\frac{\mu}{2} t'}I_{R}(t')dt'\bigg]\bigg\}, \label{DampIF}
\end{multline}
where
\be
&&I_{I}(t')= \int_{0}^{t'} \alpha_I(t',t'')\e{-\frac{\mu}{2}t''} f_{+}(t'')dt'',\\
&&I_{R}(t')= \int_{0}^{t'} \alpha_R(t',t'')\e{-\frac{\mu}{2}t''} f_{-}(t'')dt'',
\ee
are the convolution integrals containing the imaginary and real memory kernels $\alpha_I(t',t'')$ and $\alpha_R(t',t'')$
\be
    \alpha_I(t',t'')=-i\sum_k\frac{C_k^2}{2m\omega_k}\sin(\omega_k(t'-t''))
\ee
and
\be
\alpha_R(t',t'') &=& \sum_k \frac{C_k^2}{2m\omega_k}\coth(\frac{\hbar \omega_k}{2}\beta)\nonumber \\
&\times& \bigg(\cos(\omega_k(t'-t''))+\frac{\mu_0}{\omega_k}\sin(\omega_k(t'+t'')) \nonumber \\
&+&\frac{\mu_0^2}{\omega_k^2}\sin(\omega_k t')\sin(\omega_k t'') \bigg)
\ee
where
\begin{equation}
     \omega_k^2 \equiv \omega_{0_k}^2-\frac{\mu^2}{4} > 0.
\end{equation}

In order to extend to a continuum of oscillators, we define the Ohmic spectral density $J(\omega)$ 
\begin{equation}
    \omega J(\omega)= \left\{
                \begin{array}{cc}
                  2 m \e{\mu t} \eta \omega^2/\pi, & \omega<\Omega 
                  \\ 
                  0, &\omega>\Omega
                \end{array}
                \right.
\end{equation}
where we have ensured that we are consistent with the exponentially increasing mass assumption of the Caldirola-Kanai model. Using this and the high temperature limit $kT \gg \hbar\omega_k$ we obtain
\begin{equation}\label{eq:alphaR}
    \hbar \alpha_R(t',t'') \approx 2\eta kT \e{\mu t'}\bigg(\delta(t'-t'') +\frac{\mu}{2}+ \frac{\mu^2}{2}t''\bigg)
\end{equation}
and
\begin{equation}
    \alpha_I(t',t'') = i \eta\e{\mu t'} \frac{\dint}{\dint(t'-t'')} \delta(t'-t'').
\end{equation}

With these, the influence functional defined in Eq. (\ref{DampIF}) can be split into the following complex and real components, $\mathcal{F}[x,y] =\mathcal{F}_C[x,y] \; \mathcal{F}_R[x,y]$:

\be
~&&\hspace{-0.5cm}\mathcal{F}_C[x,y] = \nonumber \\ &&\hspace{-0.5cm}\exp\bigg\{\frac{iM\gamma}{\hbar}\bigg(\!-\!\int^t_0\!\left(x(t') - y(t')\right)\left(\dot{x}(t')+\dot{y}(t')\right)dt' \nonumber \\
&& \qquad \qquad \quad \quad \; \; +\mu\int^t_0\left(x^2(t')-y^2(t')\right)dt' \bigg) \bigg\} \label{FI}
\ee
and
\be
&&\mathcal{F}_R[x,y] = \nonumber \\ 
&&\exp\bigg\{-\frac{2M\gamma kT}{\hbar^2}\bigg(\int^t_0 \left(x(t')-y(t')\right)^2dt' \nonumber \\
&&+\int^t_0\!\!\int_0^{t'}\!\!\!\!\!\left(x(t')-y(t')\right)\e{\mu t'}\!\left(\frac{\mu}{2}+\frac{\mu^2}{2}t''\!\right)\e{-\frac{\mu}{2}(t'\!-t'')}\nonumber \\
&&\qquad \qquad \times(x(t'')-y(t''))\e{-\mu t''}\bigg)dt'' dt'.\bigg)\bigg\}, \label{FR}
\ee
where we have introduced the relaxation rate $\gamma=\eta/2M$, and the zero point energy divergent shift due to the interaction with the environment is reabsorbed by defining the renormalised potential of the system of interest, $V_R(x)$, as in \cite{Caldeira83}.

Eqs.(\ref{FI}) and (\ref{FR}) generalize the standard quantum Brownian motion case \cite{Caldeira83} for $\mu \neq 0$. However, while the second term of Eq. (\ref{FI}) describes a further harmonic modification of the bare potential of the system of interest, which as we shall see is responsible for interesting dynamical behaviours, the $\mu$ and $\mu^2$-proportional terms in Eq. (\ref{FR}) do not contribute to the master equation of the system. This can be seen by using the standard procedure as in \cite{Caldeira83}, and propagating the density operator from time $t$ to time $t+\varepsilon$ according to Eq. (\ref{rhoprop}). We extend Feynman's infinitesimal method \cite{Feynman65} to evaluate the time integrals in Eqs.(\ref{FI}) and (\ref{FR}) by using:
\begin{align}\label{eq:FeynamnInfinitesimal}
    S &= \int^{t+\epsilon}_t L(\dot{x}(t'), x(t'), t')dt' \nonumber \\
    &\approx \varepsilon L\bigg(\frac{x-x'}{\varepsilon}, \frac{x+x'}{2}, \frac{(t+\varepsilon)+t}{2}\bigg),
\end{align}
where $x(t)=x'$, $x(t+\varepsilon)=x$, $y(t)=y'$ and $y(t+\varepsilon)=y$. Applying Eq. \eqref{eq:FeynamnInfinitesimal} twice in Eq. \eqref{FR}, leaves a term of order $\varepsilon^2$ and therefore disregarded in the derived master equation.

Collecting the terms of order $\varepsilon$ returns the following master equation:
\be
\hspace{-0.5cm}\pdv{\tilde{\rho}}{t} &=& \frac{i\hbar }{2M}\bigg(\pdv[2]{\tilde{\rho}}{x}-\pdv[2]{\tilde{\rho}}{y}\bigg) -\frac{i}{\hbar} (V_R(x) - V_R(y))\tilde{\rho} \nonumber \\
&-&\gamma(x-y)\bigg(\pdv{\tilde{\rho}}{x} -\pdv{\tilde{\rho}}{y}\bigg)-\frac{2M\gamma kT}{\hbar^2}(x-y)^2\tilde{\rho}\nonumber \\
&+&\frac{iM\gamma}{\hbar}\mu (x^2-y^2)\tilde{\rho}\ . \label{eq:DampedME}
\ee

The first two terms in this master equation are found in the standard von Neumann equation, and the third and fourth terms are known as relaxation and decoherence terms, respectively \cite{Caldeira83,Zurek02}. However, because of the damping of the bath oscillators, a new term [$iM\gamma \mu (x^2-y^2)\tilde{\rho}/\hbar$] now appears, which can be subsumed into the second term in Eq. \eqref{eq:DampedME} as an additional potential to give a full potential, $V_F$,
\begin{equation}
    V_R(x) \rightarrow V_F(x) = V_R(x) + V_D(x), \label{eq:fullpotential}
\end{equation}
where
\begin{equation}
    V_D(x) = - \frac{1}{2} M \gamma \mu x^2, \label{eq:dampedpotential}
\end{equation}
is an inverted harmonic potential that modifies further the (renormalized) potential of the system of interest \cite{Tokieda21,Baskoutas94}. We can therefore rewrite the master equation in the familiar form of \cite{Caldeira83,Zurek02}, but with the modified potential $V_F$:
\be
\hspace{-0.5cm}\pdv{\tilde{\rho}}{t} &=& \frac{i\hbar }{2M}\bigg(\pdv[2]{\tilde{\rho}}{x}-\pdv[2]{\tilde{\rho}}{y}\bigg) -\frac{i}{\hbar} (V_F(x) - V_F(y))\tilde{\rho} \nonumber \\
&-&\gamma(x-y)\bigg(\pdv{\tilde{\rho}}{x} -\pdv{\tilde{\rho}}{y}\bigg)-\frac{2M\gamma kT}{\hbar^2}(x-y)^2\tilde{\rho}\ . \label{eq:DampedME2}
\ee

In summary, the effect of the damping on the environmental degrees of freedom manifests itself as a modification of the potential of the system of interest, without directly affecting the relaxation and decoherence terms in the master equation. However, we emphasize that this modification carries information from the environment into the system of interest, through the combined dependency on the product of the two parameters: the relaxation rate $\gamma$ and the damping coefficient $\mu$ in Eq. \eqref{eq:dampedpotential}. As a result, the potential can be tuned using both the environment's damping $\mu$ and system's relaxation $\gamma$. Changing $\mu$ will only have the consequence of changing the shape of the potential, whereas changing $\gamma$ will also affect the relaxation and decoherence terms in the master equation. Moreover, the combined dependency on $\gamma\mu$ in Eq. \eqref{eq:dampedpotential} suggests that it is also possible to keep the full potential, Eq. \eqref{eq:fullpotential}, fixed under rescaling of $\mu$ by adequately rescaling $\gamma$, which will have a direct impact on the relaxational and decoherence properties of the system. In the next section we will explore the effect of changing these variables independently of each other, and in conjunction with each other, so as to keep the full potential unchanged under different environmental dampings.

\section{Analysis of Dynamics}\label{sec:Dynamics}

In our analysis of the damped model, Eq. \eqref{eq:DampedME2}, we consider the effect of $V_D(x)$ on the dynamics of the system of interest. We choose the renormalised potential $V_R(x)$, to have a simple quartic (symmetric double well) form in which  the zero point energy divergent shift due to the interaction of the system with the environment is incorporated into a renormalised frequency, $\omega_R$,
\begin{equation}
     V_R(x) = \frac{M\omega_R^2}{2x_0^2}(x-x_0)^2(x+x_0)^2\ .\label{VR}
\end{equation}

\begin{figure}
    \centering
    \includegraphics[width=80mm]{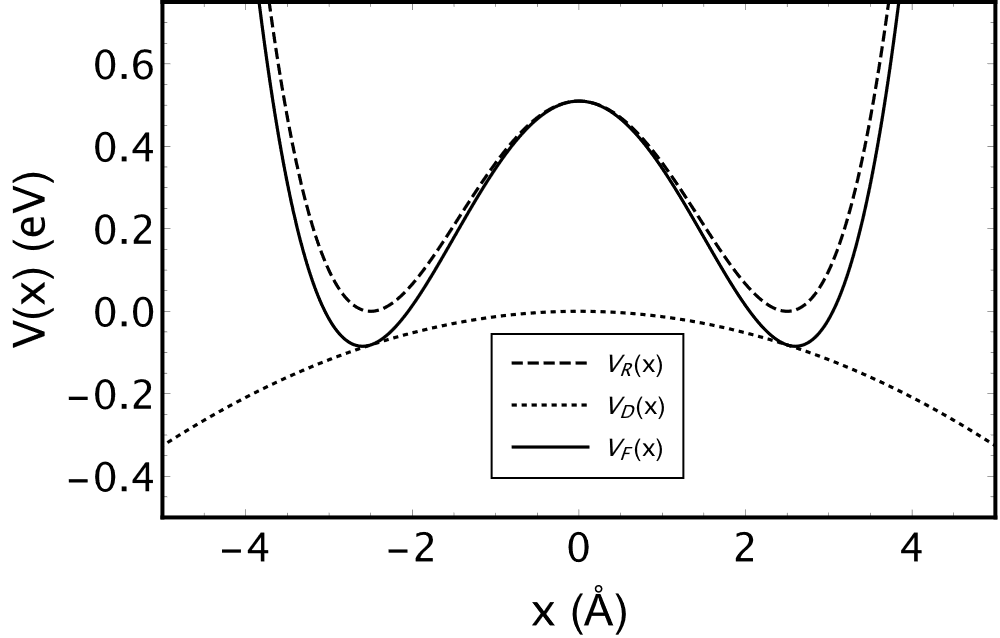}
    \caption{A plot of the three potentials:  $V_R(x)$ is the renormalised potential in the undamped master equation, $V_D(x)$ is the inverted harmonic potential due to the damped environment and $V_F(x)$, the full potential, is their sum.}
    \label{fig:potentials}
\end{figure}

A plot of all three potentials, $V_R(x)$, $V_D(x)$ and $V_F(x)$, is shown in Fig. (\ref{fig:potentials}) with the following choice of parameters: $\omega_R = 0.025$ fs$^{-1}$, $M = 938$ MeV/c$^2$, $\gamma = 2.5 \times 10^{-4}$ fs$^{-1}$, $\mu = 1$ fs$^{-1}$ and $x_0 = 2.5$\AA. We see that, compared to $V_R(x)$, both the well separation and barrier height of $V_F(x)$ increase due to $V_D(x)$. 

\subsection{Decoherence dynamics}\label{sec:coherence}

To study the effect of a damped environment on the decoherence rate, we choose the initial state of the system of interest to be a double-Gaussian with its peaks centred in the wells of the potential $V_R$, with parameter values defined as in the previous section. This choice gives us a potential barrier high enough to maintain the localisation of the two peaks over the decoherence timescale. The reduced density matrix, $\tilde{\rho}(x,y,t)$, is then evolved in time according to Eq. (\ref{eq:DampedME2}) and the rate of decoherence measured through the $l_1$-norm of coherence, $C(t)$ \cite{Baumgratz14}, by integrating over the absolute value of the off-diagonal peak where
\begin{equation}
     C(t) = \int^{L}_{0}\int^{0}_{-L} |\Tilde{\rho}(x,y,t)| ~dx~dy, \label{l1norm}
\end{equation}
and $L$ is chosen to be large enough to include the entirety of one of the off-diagonal peaks.

We compare the rates of decoherence for different values of the damping parameter $\mu$, and the relaxation rate $\gamma$, relative to the undamped case 
\begin{equation}
    C_R(t) = \frac{C_{\textrm{damped}}(t)}{C_{\textrm{undamped}}(t)}, \label{eq:Relative Coherence}
\end{equation}
where $C_{\textrm{undamped}}$ is the coherence for $\mu=0$ and corresponds to that of the standard Caldeira-Leggett model.

We start by fixing the value of $\gamma$ and varying the value of $\mu$. The range of $\mu$ values is chosen such that decoherence does not happen more quickly than the damping of the bath oscillators.

\begin{figure}[t]
    \centering
    \includegraphics[width=82mm]{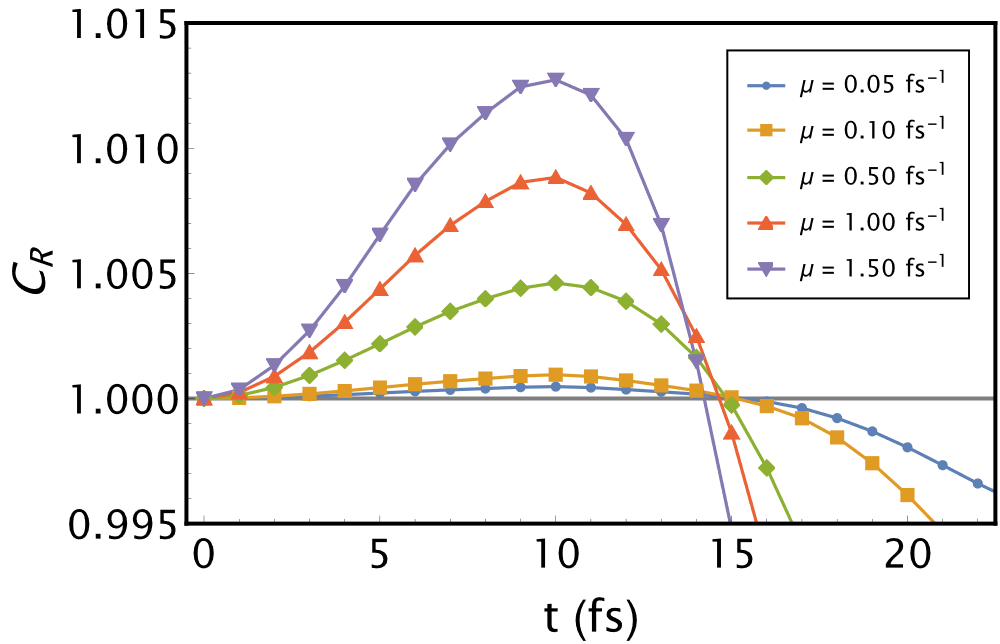}
    \caption{Time evolution of relative coherence (damped versus undamped) for a range of damping parameters, $\mu$, and a fixed dissipation parameter, $\gamma = 2.5 \times 10^{-4}$ fs$^{-1}$. Data points joined for clarity.}
    \label{fig:mu decoherence}
\end{figure}

In Fig. \ref{fig:mu decoherence} we show the behaviour of $C_R(t)$, obtained for different values of the damping coefficient $\mu$. Fig. (\ref{fig:mu decoherence}) shows that there are two distinct regimes. Initially, for $t\lesssim10$ fs, decoherence in the damped case is slower than in the undamped case, with the slowing down more pronounced the larger the value of $\mu$. Subsequently, for $t\gtrsim10$ fs, decoherence appears to become faster, and the decoherence process in the damped situation eventually becomes more efficient than in the undamped case. 

We interpret these dynamics by observing that the adopted initial condition sets the Gaussian peaks in the centres of the wells of the renormalized potential $V_R$, not of $V_F$. As a consequence, the Gaussian peaks will feel the repulsive force of the wider potential barrier, and, over time, approach their equilibrium positions, within $V_F$.

\begin{figure}[t]
    \centering
    \includegraphics[width=82mm]{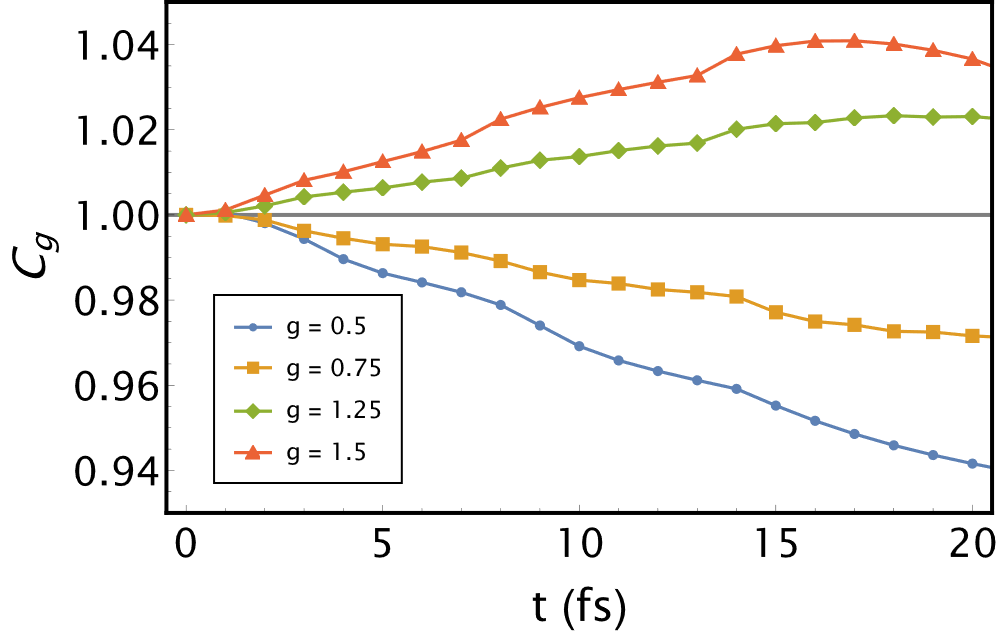}
    \caption{Time evolution of $C_g$ in the Caldeira-Leggett model (undamped environment) under the influence of a range of linear potentials of gradient $g$.}
    \label{fig:potential decoherence comparison}
\end{figure}

While doing so the system will experience different rates of decoherence. In particular the peaks are exposed initially to the gradient of $V_F$, which is steeper than that of $V_R$ at that value of $x$. We argue that it is this feature that is responsible for $C_R>1$ for $t\lesssim10$ fs. To demonstrate this, we examine the rate of decoherence for our double-Gaussian while sitting on a linear potential with various constant slopes. In Fig. (\ref{fig:potential decoherence comparison}), it is clear that as the potential slope increases, the rate of decoherence slows down. Here, we plot $C_g$, the ratio of the coherence of a double-Gaussian on a linear potential of slope $g$, against that on the reference slope $g = 1$. The reference slope has a gradient equal to that of the full potential, $V_F$, at $x = 2.5$ \AA.
 
However, after $t\gtrsim10$ fs, the Gaussians, having been pushed apart, will be far enough apart for the decoherence term in Eq. (\ref{eq:DampedME}) to dominate. In this regime, the decoherence in the full potential is faster due to the larger effect of the $(x-y)^2$ factor in the decoherence terms on the off-diagonal peaks.


\begin{figure}
    \centering
    \includegraphics[width=80mm]{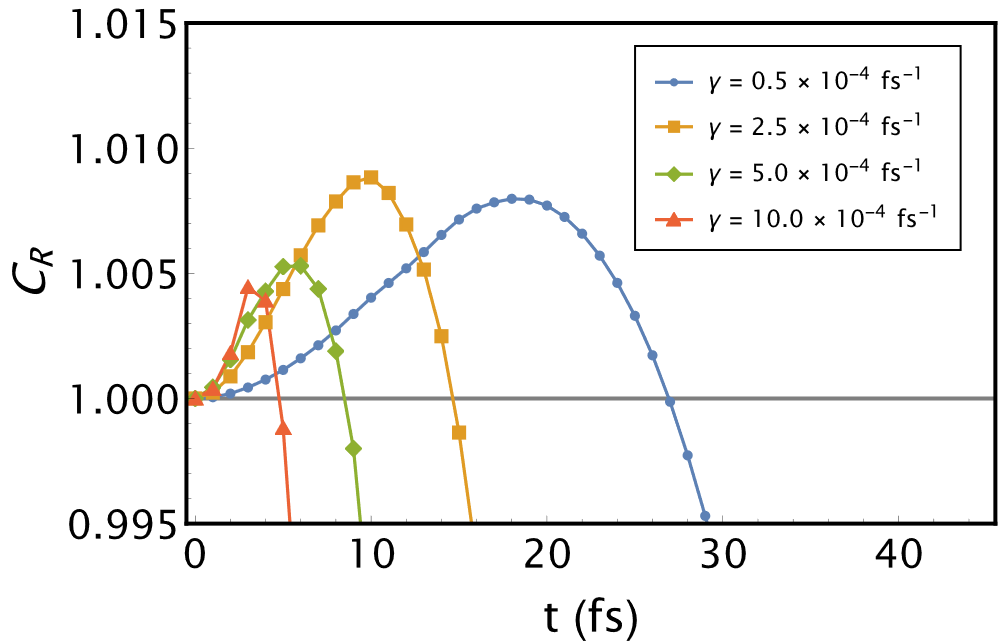}
    \caption{Time evolution of relative coherence (damped versus undamped) for a range of values of $\gamma$, and a fixed damping parameter, $\mu = 1$ fs$^{-1}$. Data points joined for clarity.}
     \label{fig:gamma decoherence}
\end{figure}

Next we fix the value of the damping parameter, $\mu$, and examine the relative coherence (for damped versus undamped environments) for various values of $\gamma$. The results are shown in Fig. (\ref{fig:gamma decoherence}). Here, we again see the behaviour in the two regimes: at short timescales the decoherence is slower in the damped environment, but becomes faster once the peaks have separated and the decoherence term in the master equation begins to dominate. Of interest here is that this switch over from slower decoherence to faster decoherence relative to the undamped case happens at progressively earlier times as the coupling between system and environment becomes stronger, suggesting that for stronger coupling, the $\gamma$-dependent decoherence term quickly dominates over the effect of the damping-induced modified barrier that delays decoherence.

\subsection{Well-transfer dynamics}\label{sec:well-transfer}

We now consider the effects of a damped environment on the dynamics of a particle initially on the right side of the potential barrier within the full potential, $V_F(x)$, but which can transfer across to the left well either by quantum tunnelling or classical over-the-barrier hopping, which together we refer to as well-transfer. We consider an initial state that is a single Gaussian centred at the minimum of the right hand well. We choose a value of $\omega_R = 0.005$ fs$^{-1}$ to give a much smaller barrier, Eq. (\ref{VR}), than we had in the previous section in order to allow for a significant transfer rate to the other well. The well-transfer probability is determined by the probability of finding the system in the left well,
\begin{equation}
    P(t) = \int^0_{-L} \tilde{\rho}(x,x,t) ~dx. \label{transferprob}
\end{equation}

It is difficult to predict intuitively in advance what the effect of the environment will be on the dynamics of the system -- the particle transfer probability in this case. On the one hand, continuous `measurement' by an environment should continuously collapse the system back to its initial state: a quantum Zeno effect, as outlined by Misra and Sudarshan \cite{misra}. However, the environment can also act to excite (thermally activate) the system, coupling it to higher energy eigenstates of the potential so that it is {\it more likely} to transfer across \cite{Godbeer14,Godbeer15,Slocombe21}.

\begin{figure}[t]
    \centering
    \includegraphics[width=79mm]{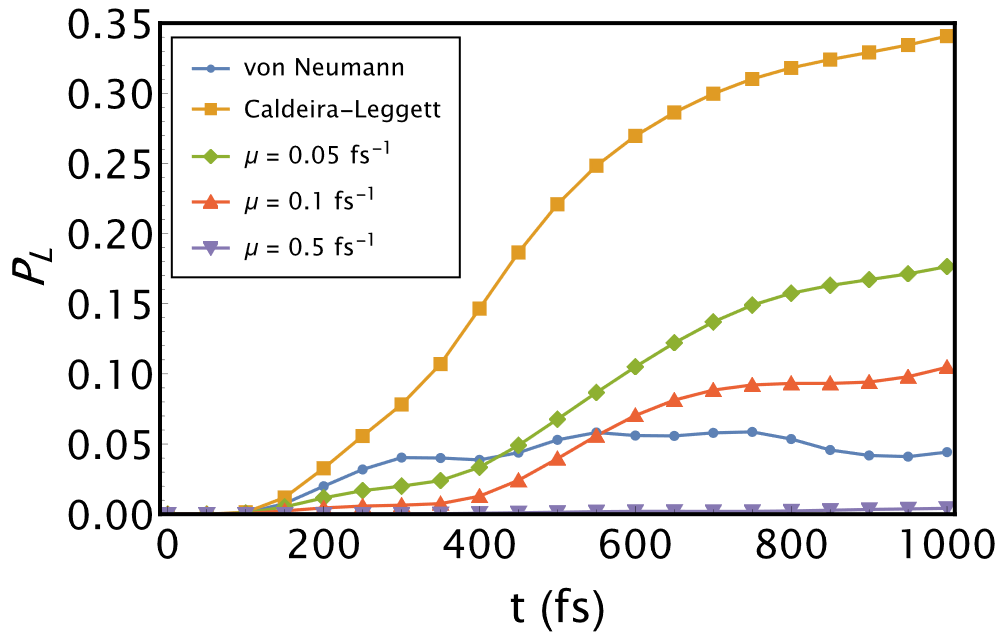}
    \caption{Time evolution of well-transfer for a range of damping parameters, $\mu$, and a fixed dissipation parameter, $\gamma = 2.5 \times 10^{-4}$ fs$^{-1}$. Data points joined for clarity.}
    \label{fig:mu tunnelling}
\end{figure}

We investigate the interplay between these two opposing effects here. Fig. (\ref{fig:mu tunnelling}) shows that over a time scale of 1000 fs the well-transfer probability in the absence of an environment (the von Neumann term only in the master equation in Eq. (\ref{eq:DampedME})) hovers around $P_L = 0.05$. By adding an undamped environment (Caldeira-Leggett model) we see a dramatic anti-Zeno effect as the system interacts with the bath oscillators, which thermally activate the particle through or over the barrier and the well-transfer probability is increased. If we then add damping to the bath oscillators, the well-transfer probability is suppressed increasingly (a Zeno effect) as the value of the damping parameter, $\mu$, is increased until there is hardly any particle-transfer at all at $\mu=0.5$ fs$^{-1}$.

\begin{figure}
    \centering
\vspace{-0.2cm}
    \includegraphics[width=78mm]{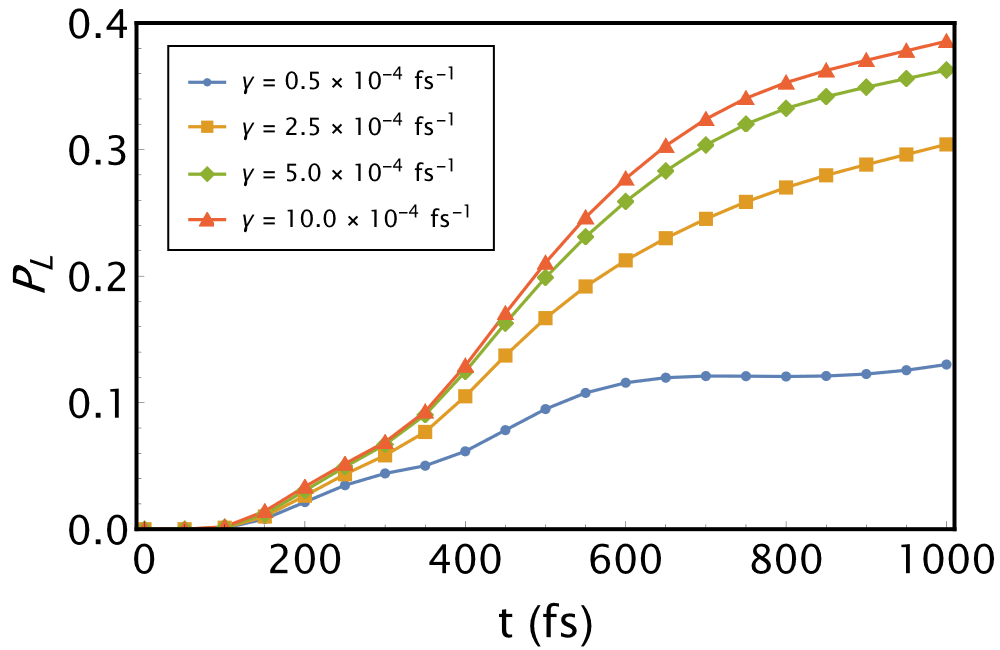}
    \caption{Time evolution of well-transfer for a range of dissipation parameters, $\gamma$, and a fixed damping parameter, $\mu = 0.01$ fs$^{-1}$. Data points joined for clarity.}
    \label{fig:gamma tunnelling D mu 0.01}
\end{figure}

In contrast, changing the relaxation rate, $\gamma$, affects the damped model in drastically different ways depending on the value of $\mu$. In the weak damping case, Fig. (\ref{fig:gamma tunnelling D mu 0.01}), we choose $\mu=0.01$ fs$^{-1}$, which is close to the Cadeira-Leggett limit of no damping at all, and note that increasing $\gamma$ increases the well-transfer probability because its effect in the master equation Eq. (\ref{eq:DampedME}) is now more strongly felt in the dissipation (3rd) term and the decoherence (4th) term.

\begin{figure}
    \centering
    \includegraphics[width=78mm]{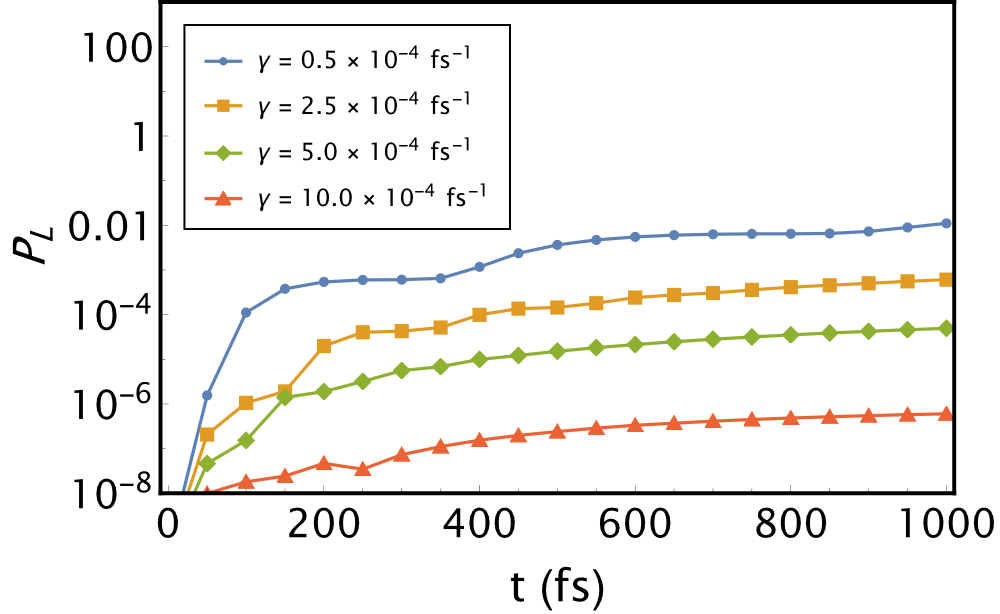}
    \caption{Time evolution of well-transfer for a range of dissipation parameters, $\gamma$, and a fixed damping parameter, $\mu = 1$ fs$^{-1}$. Data points joined for clarity.}
    \label{fig:gamma tunnelling D mu 1}
\end{figure}

However, for stronger damping ($\mu=1$ fs$^{-1}$, Fig. (\ref{fig:gamma tunnelling D mu 1})), we note the opposite effect. As $\gamma$ is increased, the well-transfer probability drops dramatically. Now, it is the final (damping) term in Eq. (\ref{eq:DampedME}) that dominates. This can be understood in terms of how it affects the modifying potential, $V_D$, Eq. (\ref{eq:dampedpotential}), which changes the shape and size of the barrier in $V_F$ in Eq. (\ref{eq:DampedME2}). The increasing barrier height and well separation with increasing $\gamma$ diminishes the probability of well-transfer (Zeno effect) more effectively than the anti-Zeno effect of the stronger coupling to the environment we saw earlier. For intermediate environment damping strengths ($0.01 < \mu < 1$), there is a transition between Zeno and anti-Zeno as $\gamma$ is increased.

\subsection{Decoherence under combined $\gamma\mu$ dependency}

In the study of open quantum systems, the reduced dynamics of the system of interest can be split into the von Neumann contribution (the unitary dynamics in the limit of an isolated system) and the additional influence of the environment due to the dissipation and decoherence terms \cite{Zurek02}. 

Thus far, we have varied each of the parameters, $\mu$ and $\gamma$, independently. In so doing, we have been changing the full potential ,$V_F$, which depends on the product $\gamma\mu$ in $V_D$. This inevitably alters the unitary dynamics of the system. However, in order to explore the effect of damping on the non-unitary terms alone, we must ensure that the potential experienced by the particle remains fixed.

For simplicity, we consider here the free particle case ($V_R = 0$, but still with the presence of $V_D$ due to environment damping) and introduce a rescaling of our parameters. We set $\gamma = \gamma_0 / a$ and $\mu = \mu_0 a$ so that $\gamma \mu = \gamma_0 \mu_0$ is constant.

\begin{figure}[t]
    \centering
    \includegraphics[width=75mm]{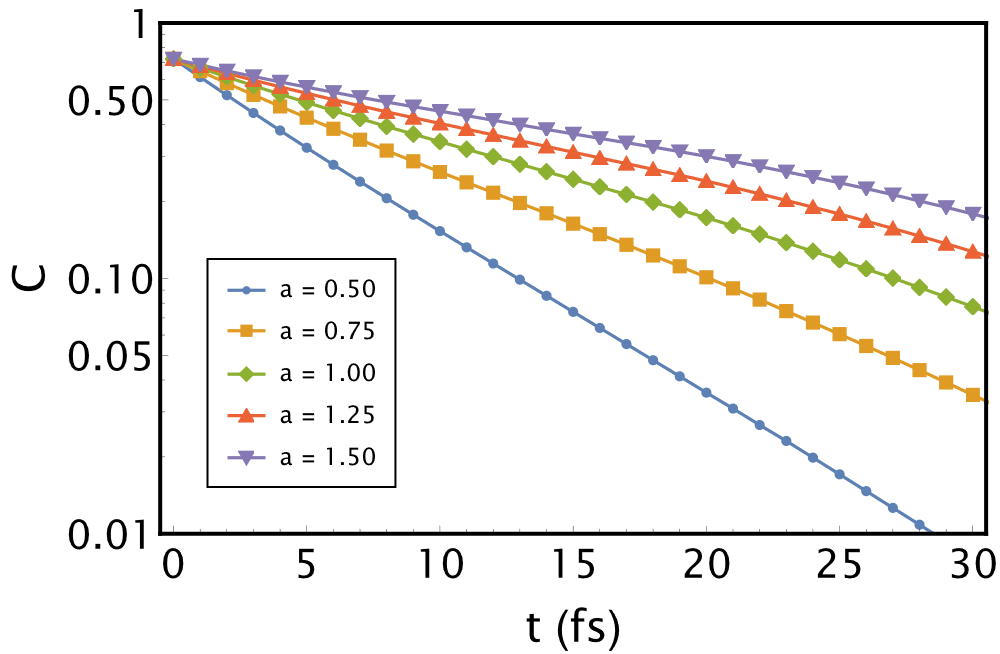}
    \caption{Time evolution of coherence for a range of rescaling factors, $a$, for $\gamma_0 = 2.5 \times 10^{-4}$ fs$^{-1}$ and $\mu_0 = 1$ fs$^{-1}$. Data points joined for clarity.}
    \label{fig:gamma-mu combined}
\end{figure}

Fig. (\ref{fig:gamma-mu combined}) shows the time evolution of the $l_1$-norm of coherence $C(t)$, defined in Eq. (\ref{l1norm}), for the case of a free particle surrounded by a bath of damped oscillators and for which the product $\gamma\mu$ is kept fixed. As the scaling parameter $a$ is increased, this turns up the strength of the damping of the bath oscillators, but simultaneously reduces the relaxation rate, which in turn results in slower decoherence. Thus, for a particle feeling a constant external potential, stronger damping of the environment results in a slower relaxation rate and weaker coupling between system and environment. 

This also tells us that taking the limit of $\gamma \rightarrow 0$, when there is no coupling between system and environment allowing the system to evolve unitarily, is equivalent to taking the limit of $\mu \rightarrow \infty$ in which the the environment oscillators are damped completely and, with no energy,  can no longer interact with the system.

\section{Conclusions} \label{sec:Conclusions}

We have presented a model of quantum Brownian motion and explored the dynamics of the reduced density matrix in the presence of a bath of damped harmonic oscillators. We describe the environmental damping by adopting the phenomenological Caldirola-Kanai model of dissipation and find that this results in an additional inverted harmonic potential that effects the unitary evolution of the system of interest.

We considered the case of a particle in a double well potential and studied the time dependence of both the decoherence rate and the well-transfer probability (at appropriate and very different timescales) by varying both the strength of the damping through the damping coefficient, $\mu$, and the relaxation rate, $\gamma$. All our findings can be interpreted as a consequence of the modification of the potential experienced by the system of interest because of the environmental damping. Initially, the superposition in the damped model experiences a steeper potential before the increase in separation of the peaks becomes significant. After an initial transient of $C_R > 1$, decoherence appears faster in the damped model because the initial coherences travel further away from the diagonal towards the more distant wells of the full potential. Similarly, well-transfer is depressed for progressively larger damping coefficients, because of increasingly deeper and more distant wells of the full potential.

It is interesting to note that in the framework introduced here, both the relaxation rate, $\gamma$, and the damping coefficient, $\mu$, enter explicitly in the unitary dynamics term of the quantum master equation, producing a coupling of dependencies between the system of interest and its environment, which is not present in the Caldeira-Leggett model. This coupling allows for a control of the decoherence process by tuning the environmental damping. Imposing the constraint that the system dynamics are  not altered upon increasing the environmental damping leads to a weakening of the coupling between the system and environment due to a reduced relaxation rate, and hence slower decoherence.

\section*{Acknowledgements} \label{sec:acknowledgements}
Lester Buxton and Marc-Thomas Russo were supported by the Leverhulme Doctoral Training Centre. Andrea Rocco and Jim Al-Khalili acknowledge support from the John Templeton Foundation, through Grant 62210. The opinions expressed in this publication are those of the author(s) and do not necessarily reflect the views of the John Templeton Foundation.

\end{document}